\documentstyle[twocolumn,pre,epsf,aps,citesort]{revtex}
\begin{document}
\draft
\title {Persistence in One-dimensional Ising Models with Parallel Dynamics}
\author{G. I. Menon\footnote{Present address: Institute for
Theoretical Physics, University of California, Santa Barbara,
CA 93106-4030, U.S.A}\footnote{Email:menon@imsc.ernet.in}
,~P. Ray\footnote{Email:ray@imsc.ernet.in}}
\address{The Institute of Mathematical Sciences, C.I.T. Campus, \\
Taramani, Chennai 600 113, India\\ {\rm and}} 
\author{P. Shukla\footnote{Email:shukla@nehu.ac.in}}
\address{Department of Physics,
North Eastern Hill University,\\
Shillong 793 022, India}
\date{\today}
\maketitle

\begin{abstract}

We study persistence in one-dimensional ferromagnetic and
anti-ferromagnetic nearest-neighbor Ising models with parallel
dynamics. The probability $P(t)$ that a given spin has not flipped up
to time $t$, when the system evolves from an initial random    
configuration, decays as $P(t) \sim 1/t^{\theta_p}$ with $\theta_p
\simeq 0.75$ numerically. A mapping to the dynamics of two decoupled
$A+A\rightarrow 0$ models yields $\theta_p = 3/4$ exactly. A finite
size scaling analysis clarifies the nature of dynamical scaling
in the distribution of persistent sites obtained under this dynamics.

\end{abstract}

Recent work on the phenomenon of ``persistence'' has revealed a novel
(and in many respects unexpected) way by which an interacting many-body
system, evolving in time, retains memory of its
initial state\cite{satya}.  Consider a one-dimensional ferromagnetic
Ising model with nearest neighbor interactions, quenched from an
initially random (infinite temperature) configuration and allowed to
relax to its global minimum energy (zero temperature) configuration, a
state with either all spins up or all spins down. Suppose the dynamics
is serial, with an attempt to update a single spin being performed at
each time step. Fix one spin and ask: What is the probability that this
spin has {\em not} flipped up to time $t$? This quantity, the
persistence probability, was first found numerically to decay as
\begin{equation}
P(t) \sim \frac{1}{t^{\theta_s}}, 
\end{equation} 
with $\theta_s$ a new,
non-trivial exponent. (The subscript refers to serial dynamics.) The
numerical results suggested $\theta_s \sim 0.37$ \cite{derrida1}; a
later analytic {\it tour de force} derived $\theta_s = 3/8$ exactly
\cite{derrida}.  Later studies have established rigorously that the
persistence exponent is a {\em new} exponent characterizing the
dynamics; it cannot be related to either the static exponents $\nu$ and
$\eta$ or the dynamical exponent $z$\cite{satya}.

This paper discusses the problem of persistence 
in the context of one-dimensional
Ising models with nearest neighbour interactions
evolving under {\em parallel} dynamics. We consider both 
ferromagnetic and anti-ferromagnetic interactions, where the 
Hamiltonian has the form
\begin{equation}
H = -J\sum_i \sigma_i\sigma_{i+1}.
\end{equation}
Without loss of generality, we take $J=1$ for ferromagnetic interactions
and $J=-1$ for the anti-ferromagnetic case. Each spin can take
values +1 (up) or -1 (down).  The zero temperature dynamics evolves a
configuration $\{\sigma(t)\}$ at time $t$ to a configuration
$\{\sigma(t+1)\}$ at time $t+1$ through the following simple rule:  For
ferromagnetic interactions, each spin at time $t+1$ assumes the value
of one of its neighboring spins at time $t$, chosen from right or left
with equal probability.  For anti-ferromagnetic interactions,
the above rule is modified in the following way: At time $t+1$,
assign the value of each spin to the {\em negative} of the value of one
of its neighboring spins, chosen from right or left with equal
probability.  Each such step in time constitutes a single Monte Carlo
step. The parallel nature of the dynamics follows from the fact that
all spins are updated together.

We have simulated parallel dynamics using the above rules on Ising
systems of linear size $L =10^2-10^6$ sites and for times $t \leq 10^5$,
applying periodic boundary conditions. We average over a fairly large
number of initial conditions, typically $10^2-10^3$ for the smaller
lattices ($L < 10^4$), starting from
configurations in which each spin is independently assigned a
value $1$ or $-1$ with equal probability.  We compute the 
standard persistence probability $P(L,t)$, defined as
the probability that the spin at a given site in a system of size $L$
has not flipped up to
time $t$, averaged over all sites and over an ensemble of initial
conditions. For $L \rightarrow \infty$ (in practice for $t \ll L^z$),
$P(L,t) \rightarrow P(t)$.

Fig. 1 shows the persistence probability $P(t)$ for 
a ferromagnetic Ising system of linear size $L=10^6$,
evolving under parallel dynamics.
$P(t)$ exhibits a power law tail
with an exponent $\theta_p \sim 0.75$. 
The behavior is  identical
for anti-ferromagnetic interactions.
For comparison, the corresponding plot for serial
dynamics is shown on the same figure; the
exponent, as advertised, is $\theta_s = 3/8$
to within numerical resolution. The
exponent $\theta_s = 3/8$ is also obtained for anti-ferromagnetic
interactions under serial dynamics.
It is thus natural to guess that
$\theta_p = 3/4 = 2\theta_s$, 
and that $\theta_p$ as well as $\theta_s$ remain unaltered when
ferromagnetic interactions are replaced by 
anti-ferromagnetic ones.

\vspace*{-3.0cm}
\begin{figure}[htb]  
\centerline{
\hbox  {
	\vspace*{-1.0cm}
	\epsfxsize=11.0cm
	\epsfbox{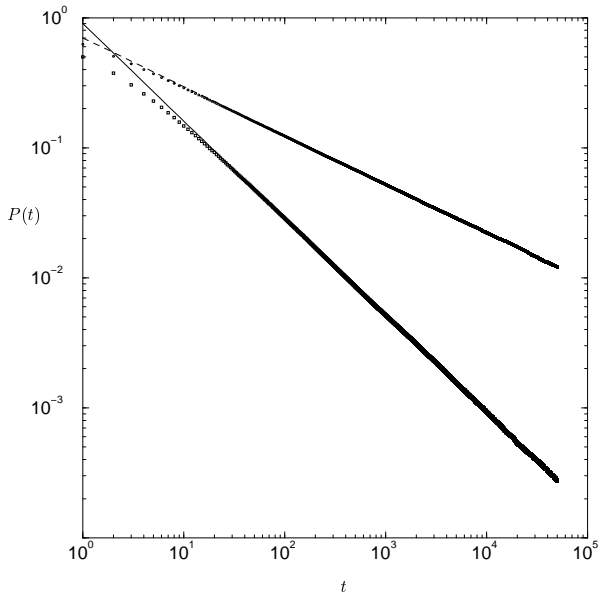}
	\hspace*{1.6cm}  
        }
	}
\vspace*{-3.5cm}
\caption
{
The persistence probability $P(t)$ in a 1d Ising model
plotted against time $t$
in a logarithmic scale. The lower curve (triangles) is for
parallel dynamics while the upper one (circles) is for
serial dynamics. The solid and the dotted lines fitted to these curves
have slopes 0.75 and 0.375 respectively. The slopes remain the same if
ferromagnetic interactions are replaced by antiferromagnetic
interactions.
}
\label{Fig1}
\end{figure}

Coarsening in ferromagnetic Ising models with serial dynamics occurs
through the motion and annihilation of domain walls separating spins
with different orientations. Spins located at domain walls have no
preference for either orientation.  A given spin flips if it is crossed
by a domain wall.  Thus, persistence in this context is equivalent to
the probability that a specified site has not been crossed by a domain
wall up to time $t$. If the domain walls in this problem are
interpreted as particles of type $A$, the following simple
reaction-diffusion scheme describes the motion and annihilation of
domain walls: $A + A \rightarrow 0$, with particles diffusing at each
time step and annihilating on contact \cite{privman}.

The problem of persistence with serial dynamics
then translates simply into
the following: Given a chosen site at time $t=0$
and an initial configuration of $A$ particles, corresponding
to domain walls in the initial configuration, what is
the probability that an $A$ particle has not crossed that
site up to time $t$? Such a redefinition of the problem
recasts the question in terms of the essential ingredients
of the dynamics, the motion and interaction of domain walls.
It is natural to look for the analog of such domain
walls in the case of parallel dynamics to understand the
conjectured relationship $\theta_p = 3/4 = 2\theta_s$.

Consider first the ferromagnetic case and divide
configurations
of spins into the following
categories:
unstable spins, implying
that they will definitely
flip in the next time step, stable spins implying
that they will not flip in the next time step,
and ``zero-field'' spins,
which may or may not flip, with either possibility
occurring with probability $1/2$. An unstable spin
$\sigma_i$ at
site $i$ has both neighbors in the state $-\sigma_i$,
while stable spins point
in the same direction as their neighbors. Zero-field
sites, on the other hand, have one neighbor pointing
up while the other points down. As a consequence,
flipping the spin at a zero-field site costs no
energy.

These zero-field sites are the analogs, for 
parallel dynamics, of domain walls
in the ferromagnetic Ising model with serial dynamics,
in a sense which we make precise below.
Stable sites belong to domains which are either
all up or all down. A spin in the bulk of
such a domain can only increase its energy if
it flips; it is thus not updated through the
zero-temperature dynamics.  Unstable sites, for the
case of ferromagnetic interactions, are
associated with an anti-ferromagnetic arrangement
of spins of the form: \ldots 1010101010 \ldots,
where the notation ``1'' indicates an up spin and ``0'' indicates a
down spin. Note that all sites interior to such a region {\em will}
flip in the next time step.  Within a region of unstable sites, spin
histories follow a two-cycle; each spin flips once in each time-step.
Sites within such regions cannot contribute to persistence, for they
cannot be persistent beyond a single time step.  It is obvious that
persistence of spin configurations at late times can only be associated
with spins which lie deep within stable regions.

It is useful for the ensuing discussion to divide the one dimensional
lattice into two inter-penetrating sublattices A and B. For example, we
may take all even-numbered sites as forming the A sublattice and
odd-numbered sites as constituting the B sublattice. The state of
sublattice A(B) at any time $t$ is determined by the state of
sublattice B(A) at $t-1$.  The initial states of sublattices A and B
are uncorrelated with each other.

We will argue below that the persistence of a site $i$ on a given
sublattice at times $t = 1,3,5 \ldots (2m+1)$ is determined by the
configuration of zero-field sites on the other sublattice, while at
times $t = 2,4,6 \ldots 2m$, it is determined by the configuration of
zero-field sites on its own sublattice.  With this in mind, it is
useful to distinguish zero-field sites on sublattice A from those on
sublattice B. Call every zero-field site on the A sublattice an A
particle and every zero-field site on the B sublattice, a B particle,
at time $t = 0$.  Fig. 2 illustrates a sequence of
configurations at succeeding instants of time through which an initial
state evolves. The letters $A$ and $B$ indicate the positions of
zero-field sites on the $A$ and $B$ sublattices.

Inspection of Fig. 2 leads to the following understanding:
Zero-field sites are located at (i) the
interface between two stable regions, in which case this interface is
composed of two adjacent zero-field sites - this is the case for the
model with serial dynamics in which case $A$ particles represent the
bound pair of zero-field sites, (ii) the interface between a stable and
an unstable region, in which case there is a single zero-field site and
(iii) the interface between two unstable regions, in which case there
is again a pair of adjacent zero-field sites. The motion of zero-field
sites corresponds to the expansion/contraction of one or the other
domain it separates.  Zero-field sites can also annihilate, leading to
the coalescence and shrinkage of domains, as the system coarsens under
the dynamics.

\vspace*{0.7cm}
\begin{figure}[htb]  
\centerline{
\hbox  {
        \epsfxsize=8.0cm
        \epsfbox{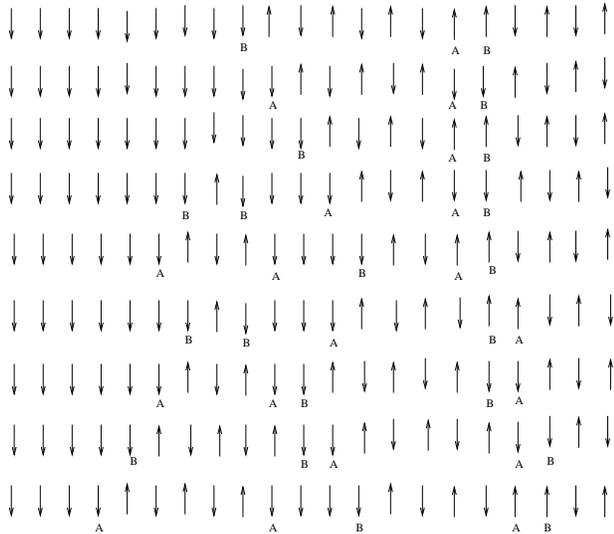}
        }
        }
\vspace*{1.0cm}
\caption
{
Time evolution of the zero-field sites
on A and B sublattices: the configuration at the earliest time
corresponds to the bottom row and each successive row is a later time
do not react with each other.  Only particles of the same kind, 
{\it i.e.} A or B,  can annihilate.
}
\label{Fig2}
\end{figure}

It is thus obvious that the non-trivial dynamics associated with the
persistence phenomenon can only be associated with the zero-field
sites, for such sites constitute the boundaries between unstable and
stable regions. It is also obvious that sites which are persistent till
time $t$ can only be associated with regions which are stable upto time
$t$, {\i.e.} regions which are stable at $t=0$ but which are not
crossed by a zero-field site upto time $t$.  The problem of persistence
in the case of parallel dynamics can thus be discussed precisely as in
the serial case, with the dynamics of the simple domain walls in the
serial case being replaced by a somewhat more complex dynamics in the
parallel case, which we deal with below.

We now establish the following crucial ingredient of this model which
enables the exact calculation of $\theta_p$, using the known result for
$\theta_s$ and the mapping onto the equivalent reaction diffusion
problem: Only particles of the same type ($A$ or $B$) can annihilate
each other.  They are transparent to particles of the other type. Thus,
the problem of the motion of zero-field sites in this model reduces to
the problem of two {\em decoupled} $A+A \rightarrow 0$ and
$B+B\rightarrow 0$ models, with the initial number of $A$'s and $B$'s
being set by the initial conditions. This is clearly evident from
Fig. 2 if we look at the configurations at even or odd time
steps and at even or odd lattice sites.  Our use of periodic boundary
conditions implies that $N(A)$ and $N(B)$ are always even, where
$N(A),N(B)$ represent the number of particles of type $A$ or $B$ at any
time.

The {\em single} conservation law of the $A+A \rightarrow 0$ model, the
conservation of particle number $N(A)$ mod $2$, is replaced by two
conservation laws in this model: $mod(N(A),2) = 0$ and  $mod(N(B),2) =
0$, where $N(A)$ and $N(B)$ are the numbers of A and B particles in any
configuration of the model.  It is useful to note the following: while
the properties conventionally computed for the reaction diffusion
scheme $A+A\rightarrow 0$ refers to sequential dynamics, these results
generalize trivially to parallel dynamics. This is a consequence of the
fact that particle moves are independent of each other in this model.
Only the reaction itself, in which two $A$ particles on the same site
annihilate each other, has different interpretations in parallel and
serial dynamics, but this ambiguity is easily removed through a simple
redefinition of the time and length-scales.

For the $A+A \rightarrow 0$ model, the density of $A$
particles decreases as \cite{wilczek}
\begin{equation}
N(A) \sim 1/t^{1/2}
\end{equation}
In the context of the Ising model with parallel dynamics, this
result implies that $N(A)$ and $N(B)$ as defined above and
at times $t, t+2, t+4 \ldots$ are conserved modulo $2$ and
decay {\em separately} as $1/t^{1/2}$, with a prefactor which depends
on the initial concentration. This result accords with 
results obtained for this model by Privman\cite{privman}; we have
checked this numerically as well.  Also note the following: If we
consider only the persistence of sites on a given sublattice, at only
even or odd time steps {\it i.e.} $t=0,2,,4 .\ldots 2m$ or $t=1,3,5
.\ldots 2m+1$, this will decay as $P(t) \sim 1/t^{3/8}$, reflecting the
fact that this dynamics maps {\em exactly} onto the $A+A \rightarrow 0$
dynamics.  This is consistent with our numerics.

We now use these results to argue the
following. Consider, for concreteness, the persistence
of a spin on the A sublattice at some time
$t$ after the quench from infinite temperature. The
persistence probability is then simply the probability
that the site in question has not been crossed
by an $A$ particle upto time $t$ {\em or}
a $B$ particle upto time $t-1$.  Since these probabilities
are independent, (the dynamics of $A$ and $B$ particles
decouples), this joint probability is simply
the product of the independent probabilities that the
site is persistent with respect to the motion of both
$A$ and $B$ particles, implying
\begin{equation}
P(t) \sim \frac{1}{t^{3/8}}\times \frac{1}{(t-1)^{3/8}}
\sim \frac{1}{t^{3/4}}
\end{equation}
yielding the persistent exponent for parallel dynamics
$\theta_{par} = 3/4$  exactly, consistent with the
numerical data.

To test the validity of the mapping onto the two non-interacting
species of particles ($A$ and $B$) outlined above, we have simulated
the associated reaction-diffusion model independently and computed,
numerically, the analog of the persistence probability for the Ising
case. This is done by computing the probability that a given site is
crossed by a particle of neither type upto time $t$; the exponent
obtained numerically tallies precisely with our result above.

How do these results generalize to the anti-ferromagnetic Ising model
with parallel dynamics? These results are unaltered as a consequence of
the following simple mapping of configurations: Replacing all spins on
one sublattice, say the A sublattice, through $\{\sigma_A\} \rightarrow
-\{\sigma_A\}$, changes the sign of the exchange interaction $J$,
mapping the problem with the new variables into the ferromagnetic
problem. This gauge symmetry relates configurations pairwise; every
update for the ferromagnetic case is an allowed update for the
anti-ferromagnet with the same weight.  Thus none of the conclusions
here are altered and the persistence exponent is independent of the
{\em sign} of the exchange interaction $J$.

Recently, there has been considerable interest in the
spatial scaling properties of persistence \cite{manoj}. Numerical work
on the 1-dimensional $A+A \rightarrow 0$ model, which
describes the Ising model with {\em serial} dynamics, shows
the existence of a non-trivial fractal structure in the
spatial distribution of persistent sites at long times.
These results can be recast in terms of a dynamical
scaling form for $P(L,t)$\cite{ray},
\begin{equation}
P(L,t) = L^{-z\theta_s} f(t/L^z),
\label{eqn}
\end{equation}
where $z$ is the dynamic exponent and $f(x) \sim x^{-\theta_s}$ for $x
<< 1$ while it is constant for large $x$.  One consequence of this
form is that persistent sites at long times
and for length scales $\ell << t^{1/z}$ constitute a fractal with
fractal dimension $d_f = d - z\theta_s = 0.25$. We have used $z = 2$,
valid for $A+A \rightarrow 0$ dynamics.

Does such structure exist for the parallel version of Ising
persistence? Our data for $P(L,t)$ are consistent with Eq.(\ref{eqn}),
with $\theta_s$ replaced by $\theta_p$ and $z \simeq 2$, as shown in
the scaling plot of Fig. 3.  This illustrates the validity of the
dynamical scaling {\it ansatz} for persistence under parallel
dynamics.  (Data collapse here is, however, inferior in comparison to
the serial case).  Using the result of the previous paragraph, this
would indicate a fractal dimension of $-0.5$,~{\it a-priori} an
unphysical result.  This result can be attributed to the fact that we
are looking at the persistence of a site under two {\em independent}
processes.  Consider the set of sites persistent with respect to the
motion of $A$ and $B$ particles separately; each will form a fractal
with the same fractal dimension $d_f$.  The intersection of these two
fractals represents those sites persistent with respect to the motion
of both $A$ and $B$ particles.  Assuming that $A$ and $B$ particles are
initially uncorrelated, the dimension of the intersection set
is then $2d_f - d = -0.5$, as above.  We conclude that persistent 
sites in the parallel dynamics version of the Ising model do {\em not}
exhibit spatial scaling of the type seen in the serial version of the
model, a result corroborated by the work of Bray and O'Donoghue
\cite{bray} and Manoj and Ray \cite{ray}.

\vspace*{-3.5cm}
\begin{figure}[htb]  
\centerline{
\hbox  {
        \epsfxsize=12.0cm
        \epsfbox{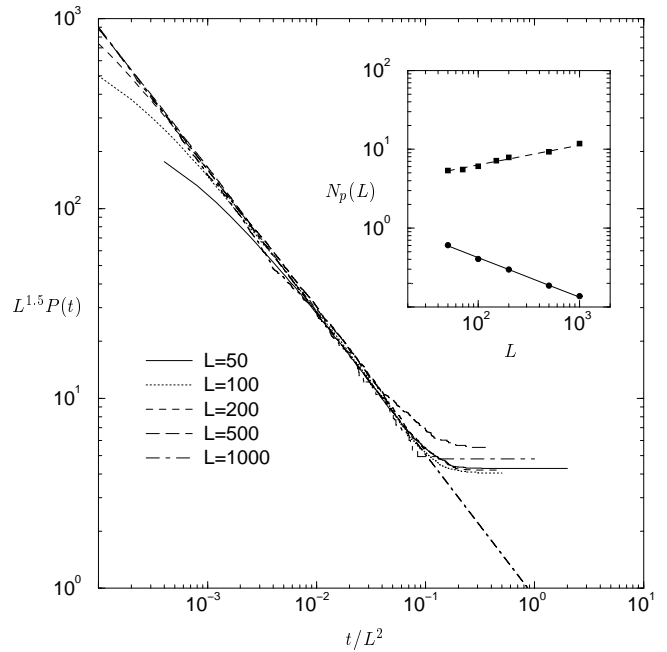}
        \hspace*{1.7cm}   
        }
        }
\vspace*{-4.0cm}
\caption
{
Plot of $P(L,t)L^{z\theta_p}$ {\it vs.} $t/L^z$, where
$P(L,t)$  is the configuration averaged density of persistent sites 
in a system of size $L$ at
time $t$, $z = 2$ and $\theta_p = 3/4$, illustrating the validity of
the dynamical scaling ansatz. The inset shows the total number of
persistent sites $N_{p}(L) = LP(L,t=\infty)$ left
in the system as $t \rightarrow
\infty$, plotted against the system size $L$ on a logarithmic scale,
both for parallel (circles) and sequential (squares) dynamics. The
straight lines fitted to these points have slopes -0.5 and
0.25 respectively (see text for discussion).
}
\label{Fig3}
\end{figure}

One implication of this result is that both the average {\em number}
and average density of persistent sites in a system of size $L$ should
{\em decay} with $L$ for parallel dynamics, in contrast to the serial
case. We have verified this numerically; see the inset to
Fig. 3.  (In contrast, the mass of a truly fractal object {\em
increases} with scale while its density decreases.)  Thus, a large
system has no persistent sites, at sufficiently long times, for most
initial conditions. The power-law tail of $P(L,t)$ is then associated
with an ever-smaller fraction, as $L$ increases, of initial states,
whose associated persistent sites survive for longer and longer times.

In conclusion, we have studied persistence in one-dimensional Ising
models with parallel dynamics. We have obtained the persistence
exponent $\theta_p = 3/4$ exactly and studied the spatio-temporal
correlations of persistent sites. These results relied on an exact
mapping to the dynamics of two {\em decoupled} $A+A\rightarrow 0$
models. It would be interesting to see if similar arguments exist and
are useful in the discussion of persistence with parallel dynamics in
other models.

We thank G. Manoj, A. Dhar, D. Dhar and M. Barma for useful
discussions.  The Associateship programme at the Institute of
Mathematical Sciences, Chennai facilitated discussions which
led to this work.  This research was supported in part by
the National Science Foundation under Grant No. PHY99-07949.

\end{document}